\documentclass[11pt]{amsart}
\usepackage{hyperref}
\font\a=cmr10  
\def \dd{{\rm d}}
\def \DD{{\rm D}}
\numberwithin{equation}{section}
\begin{document}
\title[The group aspect in the physical interpretation
of GRT] {The group aspect in the physical interpretation
of general relativity theory}%
\author{Salvatore Antoci}%
\address{Dipartimento di Fisica ``A. Volta'' and C.N.R., Pavia, Italia}%
\email{Antoci@fisicavolta.unipv.it}%
\author{Dierck-Ekkehard  Liebscher}%
\address{Astrophysikalisches Institut Potsdam, Potsdam, Deutschland}%
\email{deliebscher@aip.de}%
\maketitle {\a When, at the end of the year 1915, both Einstein
and Hilbert arrived at what were named the field equations of
general relativity, both of them thought that their fundamental
achievement entailed, inter alia, the realisation of a theory of
gravitation whose underlying group was the group of general
coordinate transformations. This group theoretical property was
believed by Einstein to be a relevant one from a physical
standpoint, because the general coordinates allowed to introduce
reference frames not limited to the inertial reference frames that
can be associated with the Minkowski coordinate systems, whose
transformation group was perceived to be restricted to the
Poincar\'e group.\par Two years later, however, Kretschmann
published a paper in which the physical relevance of the group
theoretical achievement in the general relativity of 1915 was
denied. For Kretschmann, since any theory, whatever its physical
content, can be rewritten in a generally covariant form, the group
of general coordinate transformations is physically irrelevant.
This is not the case, however, for the group of the infinitesimal
motions that bring the metric field in itself, namely, for the
Killing group. This group is physically characteristic of any
given spacetime theory, since it accounts for the local invariance
properties of the considered manifold, i.e., for its ``relativity
postulate''.\par In Kretschmann's view, the so called restricted
relativity of 1905 is the one with the relativity postulate of
largest content, because the associated Killing group coincides
with the infinitesimal Poincar\'e group, while for the most
general metric manifold of general relativity the associated
Killing group happens to contain only the identity, hence the
content of its relativity postulate is nil.\par Of course,
solutions to the field equations of general relativity whose
relativity postulate has a content that is intermediate between
the two above mentioned extremes exist too. They are the ones
generally found and investigated until now by the relativists,
since the a priori assumption of some nontrivial Killing
invariance group generally eases the finding of solutions to the
above mentioned equations. In the present chapter it is shown what
are the consequences for the physical interpretation of some of
these solutions whose relativity postulate is of intermediate
content, when Kretschmann's standpoint is consistently adhered
to.}

\newpage
\section{Introduction}\label{a}
It may seem strange that, in the year 2009, one may conceive
writing a text with the title given above. So many years have
elapsed since Einstein \cite{Einstein1915c} and Hilbert
\cite{Hilbert1915} eventually wrote the final equations of general
relativity theory, and it might be reasonable to believe that by
now the issue hinted at in the title should have been settled once
and for all. The r\^ole played by group theory in the so called
general theory of relativity should be clear beyond discussion,
and no further paper should need to be written on this
subject.\par However, this is not the case. Moreover, this problem
has emerged as early as in the year 1917, when Erich Kretschmann
defied the group theoretical assessment given by Einstein
\cite{Einstein1916}, and proposed an alternative of his own
\cite{Kretschmann1917}, whose validity in principle was to be soon
acknowledged by Einstein himself \cite{Einstein1918}. In Frank's
review \cite{Frank1917} of Kretschmann's paper one finds a short,
but precise account of the main points considered by Kretschmann.
It reads, in English translation:\par ``Einstein understands,
under his general principle of relativity, the injunction that the
laws of nature must be expressed through equations that are
covariant with respect to arbitrary coordinate transformations.
The Author shows now that any natural phenomenon obeying any law
can be described by generally covariant equations. Therefore the
existence of such equations does not express any physical
property. For instance the uniform propagation of light in a space
free from gravitation can be expressed also in a covariant way.
However, there is a representation of the same phenomena, that
admits only a more restricted group (the Lorentz transformations).
This group, that cannot be further restricted by any
representation of the phenomena, is characteristic of the system
under question. The invariance with respect to it is a physical
property of the system and, in the sense of the Author, it
represents the postulate of relativity for the corresponding
domain of phenomena.\par In Einstein's general theory of
relativity, through appropriate choice of the coordinates, the
field equations can be converted in a form that is no longer
covariant under the group of coordinate transformations. The
Author provides a series of examples of such conversions. But the
equations converted in this way in general no longer admit any
group, and in this sense Einstein's theory of general relativity
is an ``absolute theory'', while the special theory of relativity
satisfies the postulate of relativity for the Lorentz
transformations also in the sense of the Author.''\par When
reading Kretschmann's paper today, one confronts its lengthy,
sometimes obscure pages with a growing sense of admiration for the
keen physical intuition that drove its author to a right
conclusion despite his lack of the correct mathematical tools for
tackling the difficult questions that he addressed, and forced him
to try, one after another, several paths of thought that he
critically evaluated not to be fully satisfactory in one way or
another.\par In the present chapter Kretschmann's comparison
between the group theoretical assessment of the special and of the
general theory of relativity is reconsidered by availing of the
mathematical tool that is lacking in Kretschmann's work, i.e. the
group of infinitesimal Killing motions to be associated to each
theory endowed with a metric tensor. If the group properties of
both flat and curved spaces need to be compared through the same
mathematical tool, it is this group that must take the r\^ole of
what Kretschmann calls the group of invariance, the one endowed
with physical meaning, that he so many times invokes in his paper
as the one needed for properly assessing the ``relativity
postulate'' of each theory. From this recognition several relevant
consequences immediately follow for the way that must be kept when
physically interpreting the solutions to the field equations of
general relativity endowed with nontrivial groups of
invariance.\par But let us go back at present to special
relativity and to its group theoretical assessment, that has
constituted the mathematically and physically sound paradigm from
which Kretschmann has moved for building his interpretation of
general relativity as an ``absolute theory''.
\section{Finding the group of invariance in special and in
general relativity}\label{b} The reader shall forgive us if we
recall here concepts that have been perused in all the textbooks
of relativity since a long time. We need to do so for retrieving a
mathematical formulation, that may have the distinct advantage of
being applicable without change both to the special and to the
general theory of relativity.\par The Poincar\'e group of
transformations between the inertial coordinate frames
\footnote[1]{The notion reference frame acquired in the meantime a
meaning different from coordinate system, although in special
relativity both are usually intimately connected.} of special
relativity can be given in principle many representations. The
generally adopted one relies on what Landau and Lifshits
\cite{LL1970} once called ``Galilean coordinates'' $x^1=x$,
$x^2=y$, $x^3=z$, $x^4=t$, and on the Minkowski metric, expressed
with respect to these coordinates:
\begin{equation}\label{1}
\eta_{ik}=\rm{diag}(-1,-1,-1,1),
\end{equation}
that is invariant \footnote[2]{One cannot help recalling here the
ironic sentence by Felix Klein: ``Was die modernen Physiker {\it
Relativit\"atstheorie} nennen, ist die Invariantentheorie des
vierdimensionalen Raum-Zeit-Gebietes, $x$, $y$, $z$, $t$ (der
Minkowskischen ``Welt'') gegen\"uber einer bestimmten Gruppe von
Kollineationen, eben der ``Lorentzgruppe''\cite{Klein1910}.} under
the coordinate transformations of the Poincar\'e group. When this
representation is adopted, the coordinates are not just labels for
identifying events; due to the particular form of $\eta_{ik}$,
they have a direct metric reading, i.e. to each particular system
of coordinates a physically admissible reference frame, to be
built with rods, clocks and light signals, is directly associated
in one-to-one correspondence. As a consequence, by availing of
this representation, one recognizes that the Poincar\'e group,
besides being, from a mathematical standpoint, the group of
invariance of $\eta_{ik}$, is endowed with direct physical
meaning. The invariance of $\eta_{ik}$ under the Poincar\'e group
constitutes what Kretschmann once called the physically meaningful
``relativity postulate'' of the original theory of relativity.\par
However, it is quite possible, and Kretschmann was fully aware of
this
 \cite{Kretschmann1915,Kretschmann1917}, that one accounts for
special relativity by adopting a general system of curvilinear
coordinates, with the associated group of general coordinate
transformations. This move has the distinct advantage of freeing
the coordinate systems from the duplicity of function that they
play in the previous account, i.e. both providers of labels for
the identification of the events, and elements to which the
transformations of the invariance group directly apply. We do not
know whether these coordinate systems can maintain a physical
r\^ole beyond the purely topological one of identifying the
events, namely, whether reference frames can be associated with
these curvilinear coordinates too, as it was hoped for by Einstein
\cite{Einstein1916}; today, the answer to the above question is
the identification of a reference frame at a given event with the
vector base in its tangent Minkowski space-time \footnote[3]{In
this way, one obtains a field of frames which generates a
teleparallel transport with torsion instead of curvature. The
transformations of the frames form the Lorentz group at each point
separately, and event-dependent Lorentz transformations for the
field of frames. Eventually, already Einstein tried to generalize
the theory to implement the electromagnetic field in this
direction.}. However, we are sure that in general relativity the
latter coordinate systems have no relation whatsoever with the
physically relevant invariance group of the special relativity
theory \footnote[4]{The choice of the coordinate system as well as
that of the field of frames do not enter any observable
here.}.\par The adoption of curvilinear coordinates for expressing
the theory of special relativity is fundamental for acknowledging
that the restriction of the allowed coordinate systems to the ones
corresponding to the inertial frames, although very intuitive and
convenient for the calculations, is conceptually inessential. The
eventual recognition of the group of invariance of the metric in a
given theory is the true scope that we aim at, either in the
special or in the general theory of relativity, and we shall equip
ourselves with the appropriate mathematical tool. Since the
absolute differential calculus of Ricci and Levi Civita is
naturally expressed with curvilinear coordinates, these shall
constitute an appropriate choice for accomplishing our task.\par
There is a fundamental difference between the special and the
general theory of relativity, that is decisive for the very choice
of the group of invariance that we shall look at in both cases,
and for the unique mathematical tool that we shall eventually
adopt for the comparison. In special relativity, as it is evident
just because Galilean coordinates can be used in the double r\^ole
explained above, the representation of the group of invariance has
a global character, while in a nontrivial pseudo Riemannian
manifold a group of invariance of the metric, if it exists at all,
in general can be identified mathematically only in the
infinitesimal neighbourhood of each event. The several, keen but
unsuccessful attempts by Kretschmann to provide a global
identification of the invariance group through explicit analytic
or geometric procedures both in the case of special relativity as
seen in curvilinear coordinates, and in the case of general
relativity, testify the difficulty of the global problem, on which
scarce progress has occurred since Kretschmann's times.\par
Happily enough, if we investigate the invariance group of the
metric in the infinitesimal neighbourhood of each event, by
availing of the powerful tools provided by Lie and by Killing
\cite{Schouten1954} we can identify and use the algebra of the
Killing vectors that prevails in each one of these neighbourhoods,
both in the special and in the general theory of relativity. The
conceptual problem is thereby reduced to the mathematical problem
of finding the solutions of the Killing equations (\ref{A.35}) of
Appendix \ref{A} and of studying the group properties of the
infinitesimal Killing motions found in this way. As it is evident
from Appendix \ref{A}, the group of the infinitesimal Killing
motions does not deal with infinitesimal point transformations: by
its very nature, this method analyses the invariance group of the
metric under infinitesimal ``Mitschleppen'' (dragging along). This
change of objective may appear inessential for special relativity,
due to the homogeneous character of the considered manifold. In
this case, the global answer that the invariance group of the
metric is the Poincar\'e group can be reached anyway, by starting
from the infinitesimal Killing group, only through a more
complicated argument. The study of the invariance group for
infinitesimal ``Mitschleppen'', however, is the only one that is
possible in general for a pseudo Riemannian, curved manifold. The
infinitesimal Killing vector group is therefore the tool for
realizing Kretschmann's program of comparison of the invariance
groups of the metric that prevail in the special and in the
general theory of relativity respectively.\par The search for the
Killing group for both special and general relativity is
straightforward and confirms Kretschmann's objection of 1917:
while the Killing group of the metric of special relativity is the
Poincar\'e group for infinitesimal motions, for a general solution
of the field equations of general relativity the Killing group
reduces to the identity, i.e. general relativity, despite its very
name, is indeed an absolute theory.
\section{Applying Kretschmann standpoint to solutions with
intermediate relativity postulate}\label{c} Finding exact
solutions to the field equations of general relativity is a very
demanding task; no wonder then if in the decades-long search for
new solutions, since when Karl Schwarzschild discovered the
spherically symmetric, static solution that bears his name
\cite{Schwarzschild1916}, the problem has been generally eased by
limiting the search to the simpler solutions for which the Killing
groups of the metric are intermediate between the one of special
relativity and the one, endowed only with the identity, of the
most general solutions of general relativity. As a consequence,
the invariance groups of the metric fields that we can really
explore are nontrivial and, according to Kretschmann's standpoint,
intrinsic physical content is introduced a priori. Let us notice
that the idea of a particular physical content associated with a
particular nontrivial invariance group of the metric is fully in
keeping with the findings by Hilbert, Klein and with the
fundamental result by Noether \cite{Hilbert1915, Klein1917,
Noether1918} about the essential link between invariance and
conservation laws.\par We are therefore confronted with a very
interesting, but really difficult situation. The very fact that in
general relativity each particular solution of the field equations
exhibits its own particular content of the physically relevant
invariance group is a novel feature that counters our
expectations. We were prepared to search for a unique, once for
all theory of the observables of general relativity, like it
happens in special relativity, for which the Killing group is
fixed from the outset. In general relativity these observables
should behave as scalars under the group of coordinate
transformations, because tensor quantities depend on the choice of
the coordinates, which are today generally presumed to be mere
labels for identifying events, otherwise devoid of physical
meaning \footnote[5]{Scalars obtained by considering the tetrad
components of some tensor with respect to some tetrad field would
be equally devoid of physical meaning, due to the arbitrariness in
the choice of the tetrad field.}. But we do not know how to find
general exact solutions, for which these observables might display
their full structure and meaning, and even if we could find these
solutions and calculate their observables, the latter could not
have any resemblance to the observables of special relativity. In
fact, besides being invariant quantities, we know in advance that
they would obey no genuine conservation law, since the Killing
group of such general solutions would contain only the identity.
We must content ourselves, however, with the examples provided by
the particular solutions endowed with a nontrivial Killing group
which, if the Riemann tensor is nonvanishing, is different from
and endowed with less elements than the Poincar\'e group.\par Let
us explore, by availing of Kretschmann's and Noether's standpoint,
some well known solutions of general relativity, like the
Schwarzschild solution, both with the original, pondered choice of
the manifold done by Schwarzschild \cite{Schwarzschild1916}
himself and in the form, endowed with an inequivalent manifold,
accidentally introduced \footnote[6]{For a historical account on
Schwarzschild's original manifold and on the inequivalent choice
of the manifold done by Hilbert, one may consult \cite{AL2003} and
\cite{AL2006}.} by Hilbert \cite{Hilbert1917}, as well as its
Kruskal-Szekeres maximal extension \cite{Kruskal1960,
Szekeres1960}; the Kerr-Newman solution \cite{Kerr1963,
Newman1965} will be considered too. There is also a body of
literature on the so-called boost-rotation symmetric solutions
\footnote[7]{From the references on the subject let us quote here
only the solutions with nonspinning sources, reported and
investigated in \cite{Bondi1957}, \cite{BS1964}, \cite{IK1964},
\cite{Bonnor1966}, \cite{Bicak1968,BHS1983,BS1984},
\cite{Bonnor1983,Bonnor1988}, \cite{BS1989}.} that seems worth of
analysis. The perusal of these manifolds from the above mentioned
standpoint leads to disconcerting results. All these solutions,
with the exception of Schwarzschild's original manifold
\cite{Schwarzschild1916}, have one feature in common: the
manifold, on which the solution is defined, happens to be built
from the juxtaposition of submanifolds endowed with different
invariance groups of the metric, hence with different intrinsic
physical meaning, because, according to Noether
\cite{Noether1918}, the quantities that are conserved in each one
of the submanifolds are physically different.\par This peculiar
behaviour, common to the solutions mentioned above, with the
exception of Schwarzschild's original solution, is invariably due
to the presence, within the manifold, of surfaces on which the
character of one Killing vector field changes from timelike to
spacelike or vice versa, with a consequent change of the physical
meaning of the prevailing Killing group when one crosses one such
surface of junction between neighbouring submanifolds. This is a
well known behaviour, but the danger of allowing in this way for
intrinsically nonsensical, patchwork manifolds, with unrelated
physical processes, subject to unrelated conservation laws going
on severally in each of the submanifolds, has been intimated only
recently \cite{ALM2006}.\par The adoption of such composite
manifolds as models of some physical reality has occurred because
the criteria adopted for their selection have been based
exclusively on the two very important notions of local singularity
and of geodesic completeness. The two notions are deeply
intertwined in the studies that have been developed during many
years while searching for a general, invariant  and physically
satisfactory definition of singular boundary in general relativity
\footnote[8]{see for instance \cite{Geroch1968a, Geroch1968b,
Schmidt1971, GKP1972} \cite{ES1977, Thorpe1977, GLW1982,
SS1994}.}. It is not here the place for recalling them in extenso.
Suffice it to say that the notion of intrinsic, local singularity
has been associated with the divergent behaviour of the polynomial
invariants built with the metric $g_{ik}$, with the Levi-Civita
symbol $\epsilon^{iklm}$, with the Riemann tensor $R_{iklm}$ and
with its covariant derivatives, when some limit boundary is
approached along a geodesic path. A manifold is said to be
geodesically complete when its geodesics either can be defined for
any value of their affine parameter, or meet some limit boundary
where some of the above mentioned polynomial invariants diverge.
The occurrence of the latter divergence is of course an
appropriate, sufficient condition for defining a singularity
intrinsic to the manifold, and the requirement of geodesic
completeness is likely to be a geometrically and physically
correct regularity criterion for a general solution, for which the
Killing group reduces to the identity.\par When this criterion is
applied to the solutions mentioned above, for which the Killing
group does not reduce to the identity, the following assessment is
reached:
\begin{itemize}
\item{Schwarzschild's original manifold \cite{Schwarzschild1916}
is defective, because of geodesic incompleteness. No geodesic
reaching an intrinsic singularity due to the divergence of some
polynomial invariant of the Riemann tensor can be drawn on it.}
\item{Hilbert's manifold is defective due to geodesic
incompleteness too. Geod\-esics hitting an intrinsic singularity
of the previously defined kind, or emanating from it, can be
drawn, but one fails to assign to them a proper arrow of time
\footnote[9]{As required by Synge \cite{Synge1950}, a proper arrow
of time shall satisfy both the postulate of \textit{order},
according to which the affine parameter on one geodesic is always
increasing or decreasing when one goes along the geodesic in a
given sense, and the \textit{non-circuital} postulate, according
to which one cannot build, with segments of geodesics, a closed
loop on which the time arrow always points in the same sense. For
the arrow of time in Hilbert's manifold see also
\cite{Rindler2001, AL2006}.}.} \item{The Kruskal-Szekeres manifold
\cite{Kruskal1960, Szekeres1960} is geodesically complete and has
a proper arrow of time.} \item{The Kerr-Newman manifold
\cite{Kerr1963, Newman1965} lacks geodesic completion and does not
have a proper arrow of time. Both the Kerr and the Reissner -
Nordstr\"om \cite{Reissner1916, Nordstroem1918} manifolds have
been severally completed.} \item{The so-called boost-rotation
symmetric manifolds of \cite{Bondi1957}-\cite{BS1989} generally
await geodesic completion.}
\end{itemize}
When confronted with the diagram of the Kruskal-Szekeres manifold,
the perception that a consequent reasoning has eventually led us
to acknowledge the need of these four quadrants for properly
describing, in general relativity, the gravitational field of one
material particle at rest, this perception has been sufficient to
raise in some relativists the following doubt. The faultless logic
of the program of geodesic completion is of course likely to be
quite correct for a general solution to the field equations of
general relativity, for which the invariance group of the metric
contains only the identity, hence it is irrelevant. Is not it
possible that the same program may instead lead us astray when
applied to manifolds that happen to be invariantly, intrinsically
divided in submanifolds, because nontrivial and physically
different invariance groups of the metric prevail in different
parts of the complete manifold?
\par The further consideration of the infinite repetitions
occurring in the diagrams needed to perfect the program of
geodesic completion for both the Kerr and the Reissner-Nordstr\"om
solutions cannot but strengthen the doubt raised already by the
Kruskal-Szekeres manifold, and leads one to wonder whether
something similar to what occurred to Goethe's ``Zauberlehrling''
is happening here.\par If one imposes instead the condition that,
in order to be a model of some physical reality, a manifold must
not contain in its interior local, invariant, intrinsic
singularities, and must be endowed with a unique group of
invariance, the assessment of the solutions previously considered
becomes the following:
\begin{itemize}
\item{Schwarzschild's original manifold fulfills the condition.}
\item{Hilbert's manifold, that we consider here in the usual
coordinate system due to Hilbert \cite{Hilbert1917}, does not
fulfill the condition because the hypersurface orthogonal,
timelike Killing vector that can be uniquely drawn at each event
for which $r>2m$ becomes spacelike for $0<r<2m$.} \item{The
Kruskal-Szekeres manifold does not fulfill the condition for the
same reason as the one prevailing with Hilbert's manifold.}
\item{The interval of the Kerr-Newman manifold, expressed in
Boyer-Lindquist coordinates, reads:
\begin{eqnarray}\label{2}
\dd s^2 &=& -\frac{\varrho^2}{\Delta}\dd r^2 -
\varrho^2\dd\vartheta^2
-\frac{\sin^2\vartheta}{\varrho^2} ((r^2+J^2)\dd\varphi-J\dd t)^2\nonumber\\
&+&\frac{\Delta}{\varrho^2}(\dd t - J\sin^2\vartheta\dd\varphi)^2,\\
\Delta &=& r^2+J^2+Q^2-2Mr,\nonumber\\
\varrho^2 &=& r^2+J^2\cos^2\vartheta,\nonumber
\end{eqnarray} and the manifold does not fulfill the condition.
A uniform Killing group structure prevails for $r > r_0 = M +
\sqrt{M^2-J^2-Q^2}$. One can fulfill the condition by ending the
manifold there.} \item{The boost-rotation symmetric manifolds
quoted in footnote [7] do not fulfill the condition, because they
are obtained through the juxtaposition of submanifolds endowed
with physically different groups of invariance. Again, we are
confronted with a hypersurface orthogonal, timelike Killing vector
that becomes spacelike on crossing certain hypersurfaces
\cite{ALM2006}.}
\end{itemize}
\section{The singular border between submanifolds endowed with different
invariance groups}\label{d} Despite the fact that the submanifolds
into which the previously considered solutions have been divided
are invariantly defined, one might still wonder why one should
truncate manifolds that are geodesically complete, when no
singularities defined through the polynomial invariants of the
Riemann tensor occur at the borders produced in that way, and when
regular geodesics can be drawn across them. The question can be
answered by remarking that geodesics are very special worldlines,
and that the regularity of all the wordlines either crossing such
borders or lying closer and closer to them should be
investigated.\par With the manifolds considered in the previous
section, however, one does not need to accomplish such a
cumbersome program for reaching the answer. The nontrivial Killing
structure of the considered solutions allows in fact the
definition of local, invariant, intrinsic quantities besides the
just mentioned polynomial invariants, and these quantities happen
to exhibit a divergent, singular behaviour when the borders
between submanifolds endowed with different invariance groups are
approached.\par The Killing group of Schwarzschild's original
manifold \cite{Schwarzschild1916}, hence the Kill\-ing structure
of both the submanifold of Hilbert's solution \cite{Hilbert1917}
for $r>2m$, and of the left and right quadrants of the
Kruskal-Szekeres manifold \cite{Kruskal1960, Szekeres1960}, define
at each event a unique \cite{EK1964}, hypersurface-orthogonal,
timelike Killing vector $\xi_i$:
\begin{equation}\label{3}
\xi_i\xi^i>0, \ \ \xi_{i;k}+\xi_{k;i}=0, \ \ \xi_{[i}\xi_{k,l]}=0.
\end{equation}
Due to its uniqueness, and since each hypersurface orthogonal to
it is spacelike, this vector defines the unique direction of
\textit{absolute rest} in the manifold where it prevails, and
allows to build congruences of absolute rest. Let us calculate the
first curvature of one such congruence, i.e. the four-acceleration
\begin{equation}\label{4}
a^i=\frac{\DD u^i}{\dd s}\equiv\frac{\dd u^i}{\dd
s}+\Gamma^i_{kl}u^ku^l,
\end{equation}
where $\DD/\dd s$ indicates the absolute derivative, and $u^i={\dd
x^i}/{\dd s}$ is the four-velocity tangent to the chosen
congruence. From it one builds the norm
\begin{equation}\label{5}
\alpha=(-a_ia^i)^{1/2}.
\end{equation}
Due to its very definition, this local, invariant quantity is also
intrinsic to the manifold where it prevails. When Schwarzschild's
solution is written by using Hilbert's coordinates $x^1=r$,
$x^2=\vartheta$, $x^3=\phi$, $x^4=t$, its interval reads
\begin{equation}\label{6}
\dd s^2=(1-2m/r)\dd t^2-\frac{\dd r^2}{1-2m/r}
-r^2(\dd\vartheta^2+\sin^2\vartheta \dd\phi^2).
\end{equation}
We evaluate now the norm $\alpha$ of the four-acceleration along a
congruence of absolute rest for Schwarzschild's manifold, that is
accounted for in Hilbert's coordinates by (\ref{6}) with $r>2m$.
It reads
\begin{equation}\label{7}
\alpha=\left[\frac{m^2}{r^3(r-2m)}\right]^{1/2}.
\end{equation}
This local, invariant, intrinsic quantity diverges for
$r\rightarrow 2m$. It defines a singularity that one meets when
considering congruences of absolute rest closer and closer to the
inner border of Schwarzschild manifold, i.e. closer and closer to
the borders drawn in the interior of both the Hilbert and the
Kruskal-Szekeres manifolds.\par In this case the answer to the
previous question is therefore simply: the border between the
submanifolds endowed with different invariance groups of the just
examined solutions is to be considered singular from a geometric
standpoint, as soon as one does not limit the attention to the
polynomial invariants built with the Riemann tensor. As noticed
long ago \footnote[10]{before Synge \cite{Synge1966} eventually
convinced the relativists that the wise plan is to forget about
Newton's arrow and say ``gravitational field = curvature of
space-time''.} by Whittaker \cite{Whittaker1935} and by Rindler
\cite{Rindler1960}, besides the geometrical meaning, $\alpha$, and
its singularity, have an immediate physical meaning too. Let us
consider a test body of unit mass kept on a congruence of absolute
rest by a dynamometer of negligible mass; also the other end of
the dynamometer is assumed to follow a congruence of absolute
rest. According to Whittaker and Rindler, the quantity $\alpha$
then equals the strength of the gravitational pull measured by the
dynamometer \cite{ALM2001}.\par Also the left and right quadrants
of the so called boost-rotation symmetric solutions of footnote
[7] are endowed, at each event, with a unique timelike Killing
vector that is hypersurface-orthogonal with respect to a
hypersurface of spacelike character \footnote[11]{In fact, the
manifolds of these quadrants are diffeomorphic to the Weyl-Levi
Civita manifolds \cite{Weyl1917, Levi-Civita1919}, for which the
Killing group structure was examined in \cite{EK1964}.}.
Therefore, this Killing vector too uniquely defines a direction of
absolute rest, and from it a unique congruence of absolute rest is
again obtained. Since the worldlines of the material particles of
these solutions never cross the congruences of absolute rest, they
can only be interpreted as worldlines of particles in a condition
of absolute rest. Their current interpretation as worldlines of
particles executing a uniformly accelerated motion with respect to
an asymptotic reference system at spatial infinity is problematic
\cite{ALM2006}, because it relies on an approximate asymptotic
symmetry that contradicts the exact invariance group of the metric
prevailing everywhere in the submanifolds of the left and right
quadrants.\par Like it happens in the Kruskal-Szekeres manifold,
on crossing the boundaries between the left and right submanifolds
and the upper and lower submanifolds of these solutions the unique
timelike, hypersurface-orthogonal Killing vector becomes null and
then spacelike. Let us calculate the norm $\alpha$ of the
four-acceleration along congruences of absolute rest lying closer
and closer to the boundaries of the left and right submanifolds
with the upper and lower submanifolds of the solutions of footnote
[7]. We can expect \footnote[12]{in complete agreement with what
occurs to the norm $\alpha$ of the four-acceleration calculated
along a congruence of absolute rest in the left and right
quadrants of the Kruskal manifold.} that we shall find a local,
invariant, intrinsic singularity of nonpolynomial kind, associated
with the change of the invariance group that prevails there. This
is indeed the case, as it was already shown in \cite{ALM2006}, to
which the interested reader is referred for details.\par In the
Kerr-Newman solution, defined in Boyer-Lindquist coordinates by
the interval (\ref{2}), no unique, hypersurface-orthogonal,
timelike Killing vector exists. However, a singular behaviour of
$\alpha$ on approaching the boundary located at $ r=r_0 = M +
\sqrt{M^2-J^2-Q^2}$ can be invariantly proved as follows. Let
$\lambda$ and $\mu$ be two constants. The elements
\begin{equation}\label{8}
\xi^k \frac{\partial}{\partial x^k} = \lambda
\frac{\partial}{\partial t} + \mu \frac{\partial}{\partial
\varphi}
\end{equation}
of the Killing group prevailing for $r>r_0$ define invariantly a
set of orbits. The squared norm of the first curvature on these
orbits
\begin{equation}\label{9}
\alpha^2 = - g_{ij}a^ia^j = - g_{ij}
(\frac{\xi^i}{N})_{;k}\frac{\xi^k}{N}
(\frac{\xi^j}{N})_{;l}\frac{\xi^l}{N}, \ \ {\rm with}\
N=\sqrt{g_{mn}\xi^m\xi^n},
\end{equation}
contains always the factor $1/{\Delta}$ and diverges for orbits
taken closer and closer to the surface $\Delta = 0$, for which $r
= r_0 = M + \sqrt{M^2-J^2-Q^2}$. All Killing congruences defined
by (\ref{8}) are spacelike in the limit $\Delta\rightarrow 0$
except for the case given by $\mu = \lambda J/(r_0^2+J^2)\ $. This
congruence is timelike for $r > r_0 = M + \sqrt{M^2-J^2-Q^2}$ and
null in the limit $r\rightarrow r_0$. The norm of its first
curvature, i.e. the norm of its acceleration diverges in the same
limit \footnote[13]{In the case of a static metric, $J=0$, the
congruence turns out to be the hypersurface-orthogonal one.}.
Hence, in the Kerr-Newman case, at the surface $r = r_0$ a local,
invariant, intrinsic singularity is defined, despite the fact that
the polynomial invariants built with the Riemann tensor are
regular there.
\section{Conclusion}\label{e}
We all learned that the Riemann curvature is the root of all
scalar invariants that can be constructed at a certain event when
only the metric in its infinitesimal neighbourhood is known. In
the generic case (the case with trivial Killing group) this is not
questioned. However, when the Killing group of a manifold is not
trivial, its properties may produce local, intrinsic, invariant
quantities without counterpart in quantities built with the
polynomial invariants of the Riemann tensor.\par Pasting together
submanifolds endowed with nontrivial, physically different
Kill\-ing groups, besides being a move not to be recommended per
se, may produce a divergent behaviour when such invariant,
intrinsic quantities are calculated at events closer and closer to
the borders between the above mentioned submanifolds, even if the
polynomial invariants of the Riemann tensor are not divergent
there.

\appendix
\section{The infinitesimal Killing vectors}\label{A}
A very simple definition of the infinitesimal Killing vectors is
given by \cite{LL1970} and is reproduced here for the reader's
convenience. Let us consider a pseudo Riemannian manifold equipped
with two coordinate systems $x'^{i}$ and $x^{i}$ such that
\begin{equation}\label{A.31}
x'^{i}=x^{i}+\xi^{i},
\end{equation}
where $\xi^i$ is an infinitesimal four-vector. Under this
infinitesimal coordinate transformation, the components of the
metric tensor $g'^{ik}$ in terms of $g^{ik}$ read
\begin{equation}\label{A.32}
g'^{ik}(x'^{p})=\frac{\partial x'^{i}}{\partial
x^{l}}\frac{\partial x'^{k}}{\partial x^{m}}g^{lm}(x^p) \approx
g^{ik}(x^{p})+g^{im}\frac{\partial \xi^{k}}{\partial x^{m}}
+g^{km}\frac{\partial \xi^{i}}{\partial x^{m}}.
\end{equation}
The quantities in the first and in the last term of (\ref{A.32})
are calculated at the same event (apart from higher order
infinitesimals). We desire instead to compare quantities
calculated for the same coordinate value, i.e. evaluated at
neighbouring events separated by the infinitesimal vector $\xi^i$.
To this end, let us expand $g'^{ik}(x^{p}+\xi^p)$ in Taylor's
series in powers of $\xi^p$. By neglecting higher order
infinitesimal terms, we can also substitute $g^{ik}$ for $g'^{ik}$
in the term containing $\xi^i$ of the expansion truncated at the
first order term, and find:
\begin{equation}\label{A.33}
g'^{ik}(x^{p})=g^{ik}(x^{p})+g^{im}\frac{\partial
\xi^{k}}{\partial x^{m}} +g^{km}\frac{\partial \xi^{i}}{\partial
x^{m}}-\frac{\partial g^{ik}}{\partial x^m}\xi^m.
\end{equation}
But the difference
$\delta{g^{ik}(x^p)}=g'^{ik}(x^{p})-g^{ik}(x^{p})$ has tensorial
character and can be rewritten as
\begin{equation}\label{A.34}
\delta{g^{ik}(x^p)}=\xi^{i;k}+\xi^{k;i}
\end{equation}
in terms of the contravariant derivatives of $\xi^i$. When
\begin{equation}\label{A.35}
\xi^{i;k}+\xi^{k;i}=0
\end{equation}
the metric tensor $g^{ik}$ goes into itself under Lie's
``Mitschleppen'' \cite{Schouten1954}. An infinitesimal Killing
vector is a four-vector $\xi^i$ that fulfills (\ref{A.35}).

\end{document}